\documentclass{iau} 
\usepackage{graphicx}
\usepackage{comment}

\title[Tc-rich M stars] 
{ Tc-rich M stars: platypuses of low-mass star evolution}

\author[Shreeya Shetye et al.]   
{Shreeya Shetye$^{1,2}$,
 Sophie Van Eck $^2$, Alain Jorissen $^2$, Lionel Siess $^2$,
 \and Stephane Goriely $^2$}

\affiliation{$^1$ Institute of Physics, Laboratory of Astrophysics, École Polytechnique Fédérale de Lausanne (EPFL), Observatoire de Sauverny, 1290
Versoix, Switzerland\\ email: {\tt shreeya.shetye@epfl.ch} \\[\affilskip]
$^2$ Institute of Astronomy and Astrophysics (IAA), Université libre de Bruxelles (ULB), CP~226,  Boulevard  du  Triomphe,  B-1050  Bruxelles, Belgium\\}

\pubyear{2021}
\volume{366}  
\jname{The origin of outflows in evolved stars}
\editors{Leen Decin, Albert Zijlstra, Clio Gielen, eds.}
\begin{document}

\maketitle

\begin{abstract}
The technetium-rich (Tc-rich) M stars reported in the literature (\cite{little1979}; \cite{utt2013}) are puzzling objects since no isotope of technetium has a half-life longer than a few million years, and $^{99}$Tc, the longest-lived isotope along the s-process path,
is expected to be detected only in thermally-pulsing stars enriched with other s-process elements (like zirconium). Carbon should also be enriched, since it is dredged up at the same time, after each thermal pulse on the asymptotic giant branch (AGB).
However, these Tc-enriched objects are classified as M stars, meaning that they neither have any significant zirconium enhancement (otherwise they would be tagged as S-type stars) nor any large carbon overabundance (in which case they would be carbon stars). 

Here we present the first detailed chemical analysis of a Tc-rich M-type star, namely S Her. We first confirm the detection of the Tc lines, and then analyze its carbon and s-process abundances, and draw conclusions on its evolutionary status. Understanding these Tc-rich M stars
is 
an important step to constrain the threshold luminosity for the first occurrence of the third dredge-up and the composition of s-process ejecta during the very first thermal pulses on the AGB.
\end{abstract}

\firstsection 
\section{Introduction}

The resonance lines of the radioactive 
element technetium (Tc) (i.e., having no stable isotopes) were first identified in several S-type stars by \cite{merrill}. 
 The isotope $^{99}$Tc is the only one to be located along the path of the slow neutron-capture nucleosynthesis (s-process) occurring in the interior of asymptotic giant branch (AGB) stars. $^{99}$Tc has a half-life of about 2~x~10$^5$~yr. Along with carbon and other s-process elements, $^{99}$Tc is brought to the surface of AGB stars by a mixing process called the \textit{third dredge-up} (TDU; Iben \& Renzini 1983). 
The detection of Tc in some -- but not all -- S stars led to the discovery that these evolved stars come in two flavors: the Tc-rich stars (also known as `intrinsic' S stars) are genuine thermally-pulsing AGB stars, while the Tc-poor stars\footnote{More information on Tc-poor S stars can be found in \cite[Jorissen \etal\  (1988, 1993)]{jorissen88,jorissen93}, \cite[Van Eck \& Jorissen (1999, 2000b)]{SVE1999, SVE2000b}, \cite{SVE2000a}
and \cite{shetye1}.} (also known as `extrinsic' S stars) owe their s-process enhancement to a binary interaction with a former AGB companion (\cite[Iben \& Renzini 1983]{iben1983}, \cite[Jorissen \etal\ 1993]{jorissen93}). Hence, the presence of Tc is a very sensitive probe of the TDU and AGB nucleosynthesis. 

Stars experiencing the TDU are expected to show at their surface an amount of carbon and of s-process elements increasing with the number of thermal pulses and should evolve according to the spectral sequence M $\rightarrow$ MS $\rightarrow$ S $\rightarrow$ SC $\rightarrow$ C stars. The spectral classification from M to S is based on the identification in M stars of TiO bands only and of TiO and ZrO bands in S-type stars (where Zr originates from the s-process enhancement).
The Tc-rich nature of some Mira variables of spectral type M was discovered by \cite{little1979} and \cite{little1987}.  The Tc-rich M stars constitute an interesting group of stars because they show clear signatures of Tc but no presence of ZrO bands. Hence, the Tc-rich M stars could be the first objects on the AGB to undergo a TDU, where the Tc can be detected in the spectrum before enhancements of other s-process elements become measurable (\cite{stephane2000}). 

Despite their importance, a detailed spectroscopic investigation of the Tc-rich M stars is still lacking. The spectroscopic investigation of a large-sample of M stars is currently on-going, and will be published in a forthcoming paper (Shetye \etal.~in prep.). In the current work, we discuss the pilot study of the Tc-rich M star S~Her. S~Her was first identified as a Tc-rich MS star by \cite{little1987}. \cite{keenan1974} classified it as M4-7.5Se. In the current work, we confirm the presence of Tc from a high-resolution spectrum of S~Her. We then examine its s-process element abundances and report our findings in Section 3.


\section{Spectral Analysis} 


We collected high-resolution spectra of S Her 
with the HERMES spectrograph (\cite[Raskin et al. 2011]{raskin}) mounted on the 1.2m
Mercator Telescope at the Roque de Los Muchachos Observatory, La Palma (Canary Islands). HERMES spectra have a spectral resolution of R = 85 000 and a wavelength coverage 380–900 nm. The HERMES spectra of S Her used in the current analysis have a signal-to-noise ratio of 60 in the $V$ band to ensure the accurate determination of the stellar parameters and of the chemical abundances. It is important to keep in mind that the Tc-rich Mira M stars 
experience severe stellar variability, making their spectral analysis 
difficult with 1D, static model atmospheres.
According to the AAVSO\footnote{https://www.aavso.org/} International Database, S Her has a
peak-to-peak variability in the $V$ band
of roughly 5 magnitudes. 

\begin{figure*}[!htbp]     
\begin{centering}
    \mbox{\includegraphics[scale=0.3, trim={2cm 0cm 4cm 0cm}]{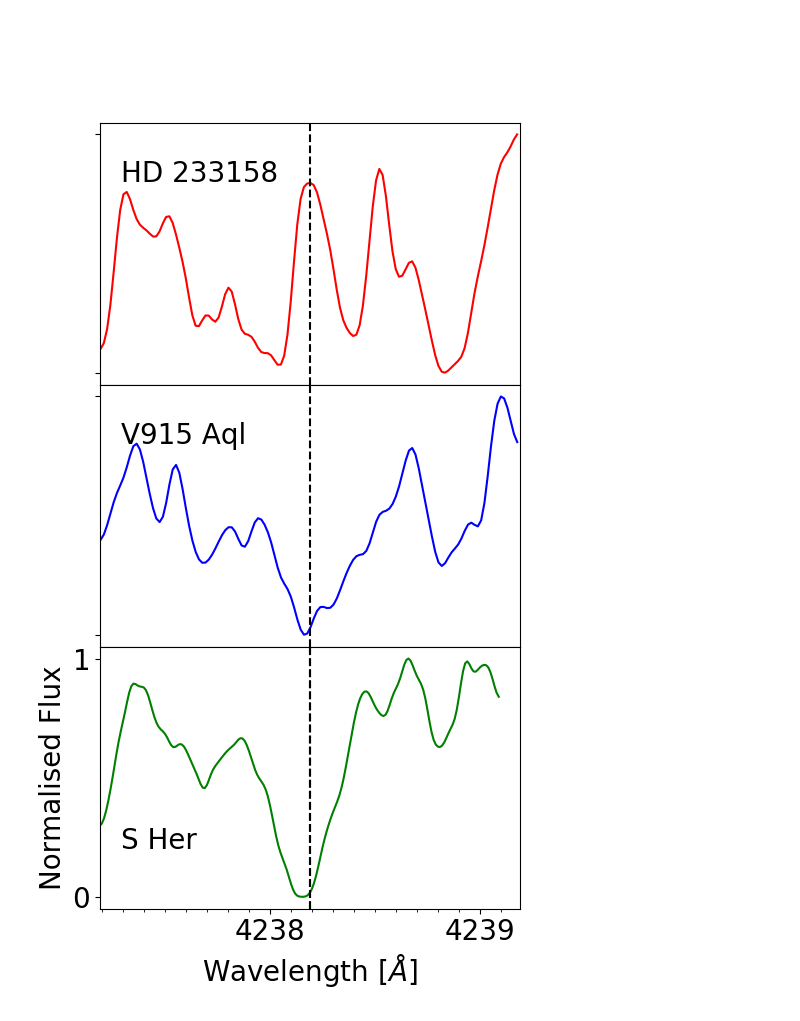}}   
    \hspace{0px}
    \mbox{\includegraphics[scale=0.3, trim={2cm 0cm 4cm 0cm}]{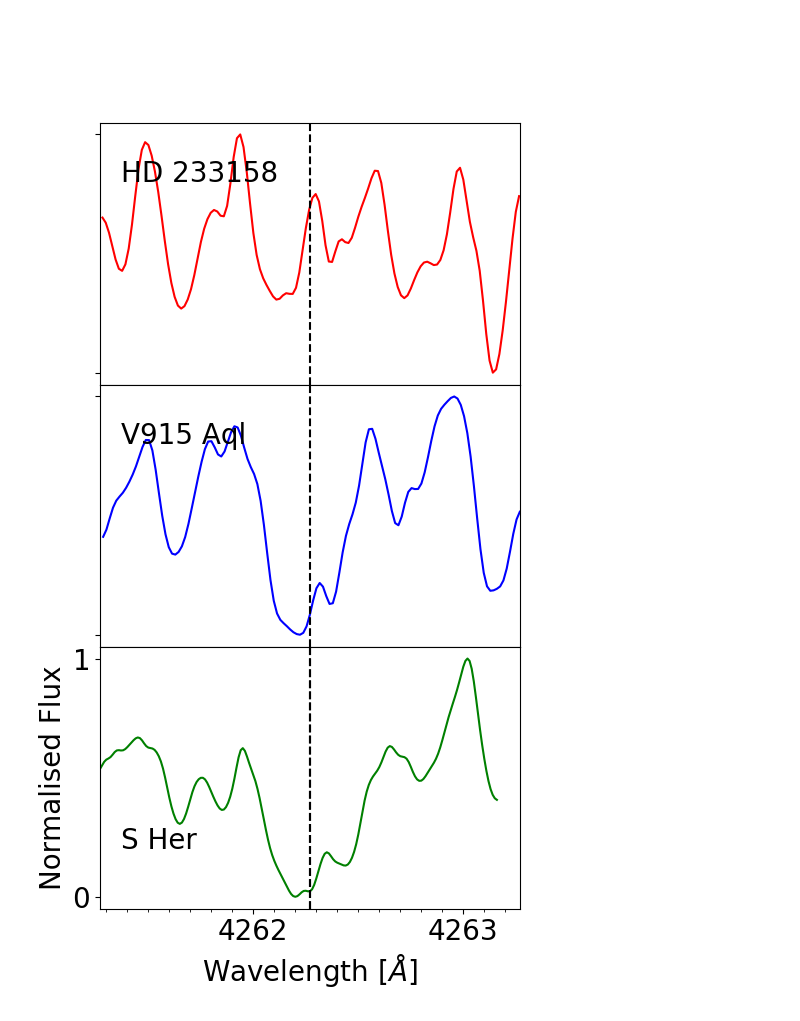}}
    \hspace{0px}
    \mbox{\includegraphics[scale=0.3, trim={2cm 0cm 4cm 0cm}]{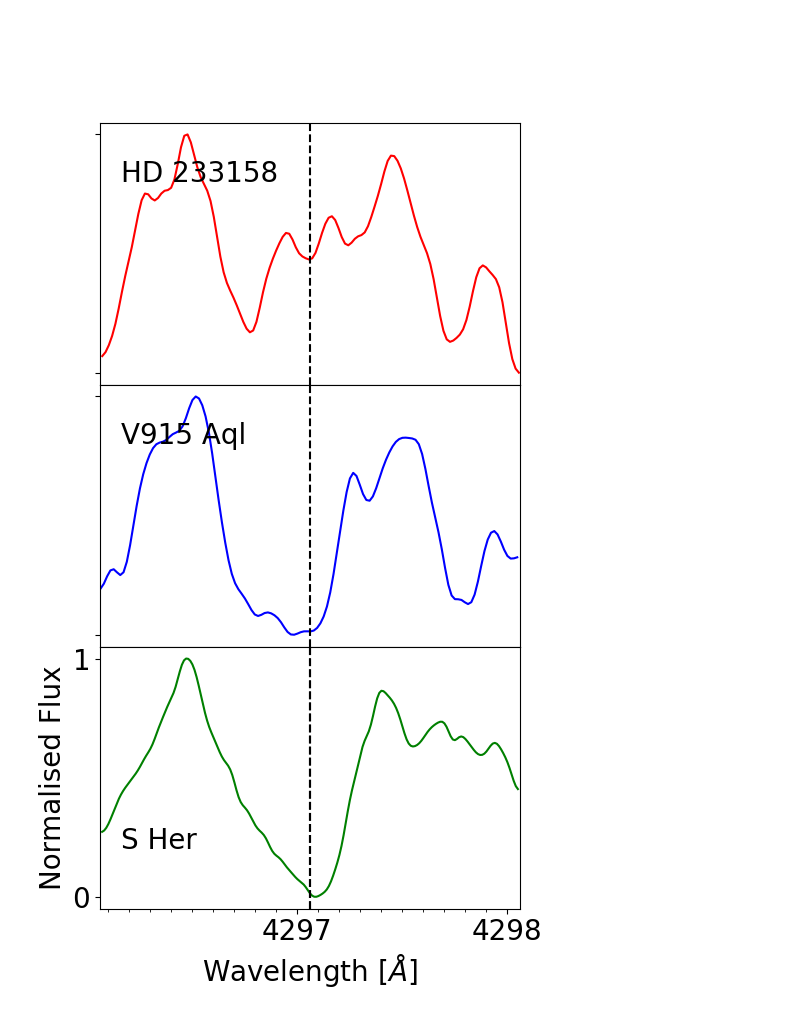}}
    \caption{\label{TcMstars}
    Spectral region around the three Tc I lines 
    (4238.19, 4262.27 and 4297.06~\AA)  for the Tc-rich M star S Her. For comparison purposes, the spectra of a Tc-poor S star (HD 233158, in red in the top  panels) and of a Tc-rich star (V915~Aql, in blue in the middle panels) from \cite[Shetye \etal\ (2018)]{S18} are also plotted. The spectra have been arbitrarily normalized and binned by a factor of 1.5 to increase the S/N ratio.}
\end{centering}
\end{figure*}

The initial step in the spectral analysis of S Her was to confirm the presence of Tc in its spectrum. We used the three Tc I resonance lines located at 4238.19, 4262.27 and 4297.06 \AA. In Figure \ref{TcMstars}, we compare the three Tc lines of S Her with an extrinsic (top panel) as well as an intrinsic (middle panel) S star from \cite{S21}. From Figure~\ref{TcMstars}, we report an unambiguous detection of Tc in the spectrum of S~Her, hence making it a good candidate for our case study of Tc-rich M stars. 

The stellar parameters of S Her are determined using the method described in \cite{shetye1}. This method was developed for the stellar-parameter derivation of S stars and was 
applied on a large sample of Tc-rich S stars (\cite[Shetye \etal\ 2019, 2020, 2021]{S19,S20,S21}). We refer the reader to \cite{shetye1} for 
details.
In summary, this method compares the high-resolution HERMES spectra to a grid of synthetic S-star MARCS model spectra (\cite[Van Eck \etal\ 2017]{SVE2017}). This grid spans a large range in effective temperatures, in surface gravities, and most importantly in the carbon/oxygen ratios and the s-process enhancement levels, the latter two being 
relevant
parameters
since they impact significantly 
thermally-pulsing AGB star spectra. Furthermore, we located S Her in a color-color diagram using the photometric indices $J-K$ and $V-K$ as described in Section 3 and Figure 5 of \cite{SVE2017}. 
This photometric estimate of the stellar parameters is consistent with the spectroscopic one. The adopted parameters are listed in Table~\ref{tab1}.


We used the atomic line lists provided by \cite[Shetye \etal\ (2018, 2021)]{shetye1,S21} for the investigation of s-process abundances in S~Her. The results 
of the abundance analysis of 
a selection of heavy elements are described in the next section.

\begin{table}
  \begin{center}
  \caption{The stellar parameters of S Her used in the current analysis.}
  \label{tab1}
 {\scriptsize
  \begin{tabular}{|l|c|c|c|c|c|}\hline 
{\bf Teff} & {\bf log g} & {\bf [Fe/H]} & {\bf C/O} & {\bf [s/Fe]} &  {\bf $L$}  \\ 
(K) & (dex) & (dex) &  & (dex) & (L$_{\odot}$)  \\
\hline
2700 & 0.0 & 0.0 & 0.5 & 1.0  & 4200  \\
 \hline
  \end{tabular}
  }
 \end{center}
\vspace{1mm}
 \scriptsize{
 {\it Notes:}\\
  $^1$ The method to derive the atmospheric parameters is described in Section 2.\\
  $^2$ The luminosity was derived using the Gaia eDR3 parallax and the same method as described in \cite[Shetye \etal\ (2018, 2021)]{shetye1,S21}.}
\end{table}

\section{Implications}

\subsection{The chemical status of S Her}

According to Figure~\ref{TcMstars}, S Her is undoubtedly a Tc-rich M star. 
The detection of Tc is 
an unambiguous evidence for the occurrence of the third dredge-up and s-process nucleosynthesis. The presence of Tc in AGB stars undergoing TDU episodes is usually accompanied by an overabundance in other s-process elements. Even 
though S~Her shows clear Tc
lines, our analysis could not detect any enrichment in other s-process elements. The match between the observed and synthetic spectra in many spectral windows is however far too bad to derive abundances. Conversely, in the wavelength regions where the match is acceptable, there are either very few s-process lines present or they are not good abundance probes given the low temperature of S~Her. In conclusion, the low temperature of S~Her and its large stellar variability make the derivation of s-process abundances extremely challenging (even almost impossible!).

\begin{figure*}
\begin{centering}
\includegraphics[scale = 0.275]{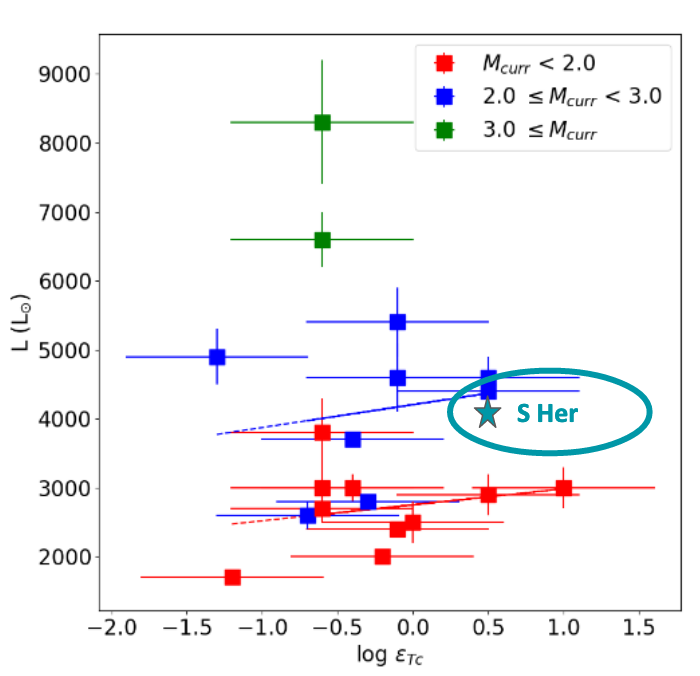}
\caption{\label{TcSHer} Figure adapted from Figure 9 of \cite{S21} to compare the stellar luminosity of S~Her and its Tc abundance with that of Tc-rich S stars of various masses, according to the color coding given in the inset. The blue and red dashed lines represent the linear least-square fit for the stars in the corresponding mass bin.}
\end{centering}
\end{figure*}

The spectral window including the Tc line regions is well reproduced by the synthetic spectra. We used the 4262.27~\AA\ Tc~I line to derive the Tc abundance. This line is the least blended line amongst the three available Tc lines (\cite[Little-Marenin \etal\ 1979]{little1979}; \cite[Shetye \etal\ 2021]{S21}). We derived an upper limit of log $\epsilon_{Tc} \sim$ +0.5 dex on the Tc abundance of S Her. In Figure \ref{TcSHer}, we compare the location of S Her in the luminosity vs Tc abundance diagram from \cite{S21}. Surprisingly, this upper limit for the Tc abundance in the M star S~Her is comparable with 
the highest Tc abundances
encountered in S-type stars
from the sample of \cite{S21}. 

\subsection{S Her and R Dor}

\begin{figure*}
\begin{centering}
\includegraphics[scale = 0.52]{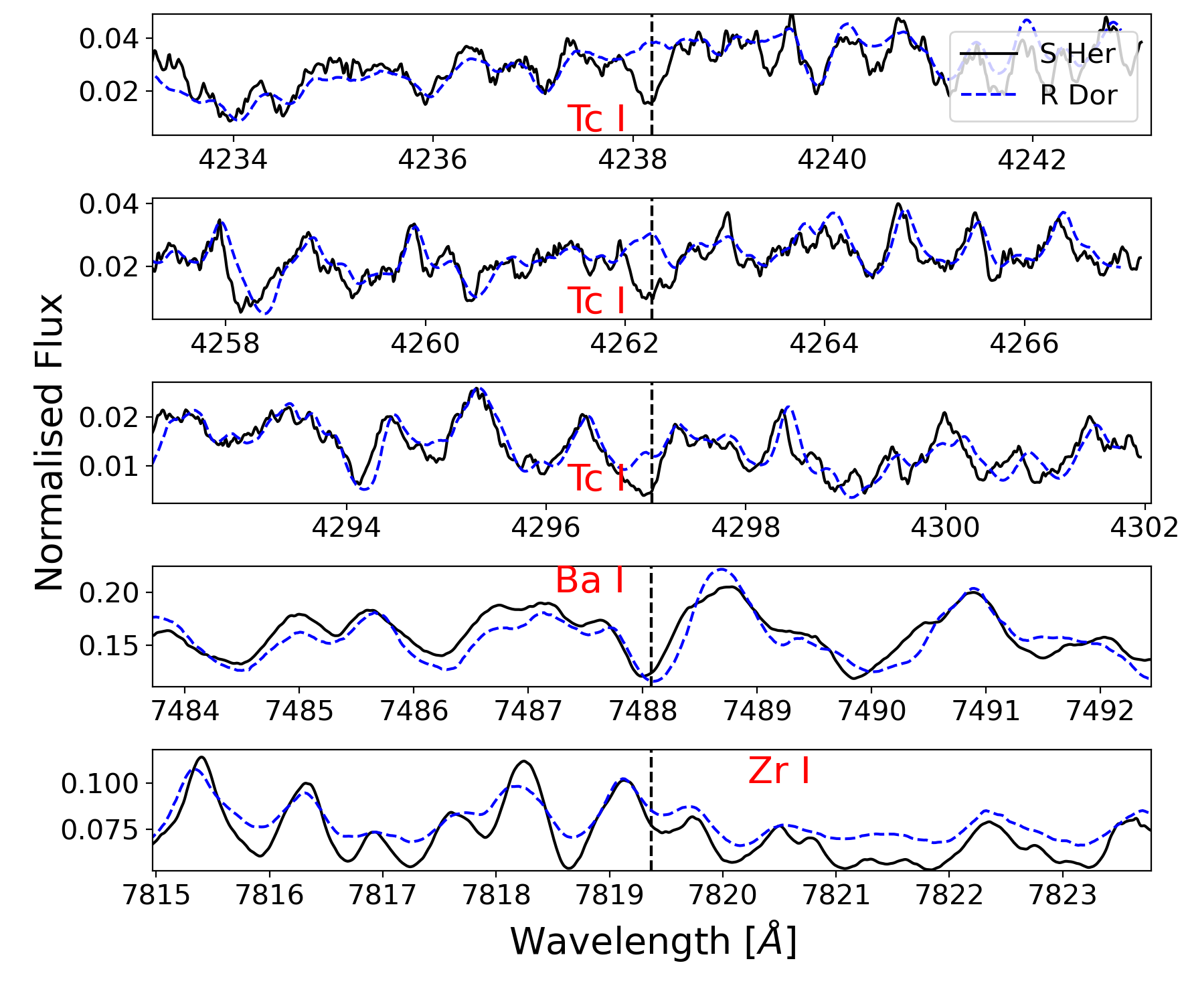}
\caption{\label{sprocesslines} Comparison of s-process lines in S Her (black, continuous line) and R Dor (blue, dashed line). From top to bottom: Tc~I line at 4238.191~\AA, Tc~I line at 4262.27~\AA, Tc~I line at 4397.06~\AA, Ba~I line at 7488.077~\AA, and Zr~I line at 7819.37~\AA. }
\end{centering}
\end{figure*}

Considering the above-mentioned difficulties faced by spectral synthesis to derive abundances
we decided to 
consider a differential analysis instead. We 
looked for an M-type giant star with properties similar 
to those of S~Her (mainly in terms of its effective temperature and variability). We found that
the stellar parameters of R~Dor, classified as M8III:e by \cite{keenan1989}, 
are very similar to those of
S~Her. The UVES\footnote{https://www.eso.org/sci/observing/tools/uvespop/interface.html} spectrum of R~Dor shows exceptional resemblance with that of S~Her, according to  Figure~\ref{sprocesslines}  which
 compares lines from Tc and other s-process elements in R~Dor and S~Her. From the first three panels of Figure~\ref{sprocesslines}, it is obvious that the spectra of R~Dor and S~Her are identical except for the Tc lines. 
As shown above, S~Her is clearly a Tc-rich M star 
while R~Dor is not Tc-rich. 
For Ba and Zr lines, the difference is not as clear (panels 4 and 5 of Figure~\ref{sprocesslines}). We also inspected some r-process element lines and found no clear difference either between R~Dor and S~Her. 

\section{Prospects}
The Tc-rich M stars 
constitute an intriguing class of AGB stars. It remains difficult to 
decide whether the non-detection of overabundances for other s-process elements (apart from Tc) in S~Her is spurious and results from the 
complexity of optical spectra of Mira variables or if Tc-rich M stars are truly AGB stars caught just after their first thermal pulse and with no s-process signature other than the presence of Tc. 
Nevertheless, Tc-rich M stars might carry clues about the luminosity of the first occurrence of the TDU and its dependence on the mass and metallicity of the stars. But such clues remain to be unfolded.


In the future, investigating the Tc-rich M stars should proceed along the following avenues:


- \textit{Going to infrared spectroscopy}\\
The optical spectra of oxygen-rich AGB stars are dominated by molecules making the atomic diagnostics quite difficult. The near-infrared regions have been successfully used for the chemical analysis of AGB stars (\cite[Smith \& Lambert 1985, 1990]{smith1,smith2}). It will be interesting to study the near-IR spectral windows, with the goal of finding some s-process lines as well as getting constraints on the carbon isotopic ratio. A combined study of optical (for Tc) and infrared spectra 
thus appears 
essential.\medskip\\

- \textit{Large-scale chemical investigation of Tc-rich M stars}\\
The chemical 
pattern of S Her 
is quite puzzling. The presence of Tc without
any other s-process overabundances might either be a special characteristic of S Her or be common among Tc-rich M stars. Hence, a large-scale study of M stars (Shetye \etal, in prep.) will 
be important to check for a possible diversity among the Tc-rich M stars. 

Furthermore, the derived chemical abundances of such a sample of Tc-rich M stars can be used for a quantitative comparison between observations and stellar evolution models. \medskip\\


- \textit{A spectroscopic investigation using non-static AGB model atmospheres}\\
The current poor agreement between the synthetic and observed spectra of Tc-rich M stars could be resolved by making use of non-static AGB models. In the future, abundance analysis using 3D hydrodynamic models should be attempted.

\section*{Acknowledgements}
\noindent LS and SG are FNRS senior research associates. SVE thanks Fondation ULB for its support. Based on observations obtained with the Mercator Telescope and the HERMES spectrograph, which is supported by the Research Foundation – Flanders (FWO), Belgium, the Research Council of KU Leuven, Belgium, the Fonds National de la Recherche Scientifique (F.R.S.-FNRS), Belgium, the Royal Observatory of Belgium, the Observatoire de Gen\`eve, Switzerland and the Th\"{u}ringer Landessternwarte Tautenburg, Germany.


\begin{thebibliography}{}
\bibitem[Goriely \& Mowlavi (2000)]{stephane2000}
 {{Goriely}, S. and {Mowlavi}, N.} 2000,
 \textit{A\&A}, 362, 599G

\bibitem[Iben \& Renzini (1983)]{iben1983}
 {{Iben}, I., Jr. and {Renzini}, A.} 1983,
 \textit{ARA\&A}, 21, 271I
 
 \bibitem[Jorissen \etal~(1988)]{jorissen88}
 {{Jorissen}, A. and {Mayor}, M.} 1988,
\textit{A\&A}, 198, 187J
 
 \bibitem[Jorissen \etal~(1993)]{jorissen93}
 {{Jorissen}, A., {Frayer}, D.~T., {Johnson}, H.~R., {Mayor}, M. \& {Smith}, V.~V.} 1993,
 \textit{A\&A}, 271, 463J

\bibitem[Keenan \etal~(1974)]{keenan1974} Keenan, P.~C., Garrison, R.~F., \& Deutsch, A.~J.\ 1974, \textit{ApJS}, 28, 271

\bibitem[Keenan \& McNeil (1989)]{keenan1989} Keenan, P.~C. \& McNeil, R.~C.\ 1989, \textit{ApJS}, 71, 245

\bibitem[Little-Marenin \& Little (1979)]{little1979}
 {{Little-Marenin}, I.~R. and {Little}, S.~J.} 1979,
 \textit{AJ}, 84, 1374L
 
 \bibitem[Little \etal~(1987)]{little1987}
 {{Little}, S. J., {Little-Marenin}, I. R., \& {Bauer}, W. H.} 1987,
 \textit{AJ}, 94, 981L
 
 \bibitem[Merrill (1952)]{merrill}
 {{Merrill}, P. W.} 1952,
 \textit{ApJ}, 116, 21M

\bibitem[Raskin \etal~(2011)]{raskin}
{{{Raskin}, G., {van Winckel}, H., {Hensberge}, H., {Jorissen}, A., {Lehmann}, H., {Waelkens}, C., {Avila}, G., {de Cuyper}, J. -P., {Degroote}, P., {Dubosson}, R., {Dumortier}, L., {Fr{\'e}mat}, Y., {Laux}, U., {Michaud}, B.,  {Morren}, J., {Perez Padilla}, J., {Pessemier}, W., {Prins}, S., {Smolders}, K., {van Eck}, S., \& {Winkler}, J.}} 2011,
\textit{A\&A}, 526A, 69R

\bibitem[Shetye \etal~(2018)]{shetye1}
{{Shetye}, S., {Van Eck}, S., {Jorissen}, A., {Van Winckel}, H., {Siess}, L., {Goriely}, S., {Escorza}, A.,  {Karinkuzhi}, D., \& {Plez}, B.} 2018,
\textit{A\&A}, 620A, 148S

\bibitem[Shetye \etal~(2019)]{S19}
{{Shetye}, S., {Goriely}, S., {Siess}, L., {Van Eck}, S., {Jorissen}, A. \& {Van Winckel}, H.} 2019,
\textit{A\&A}, 625L, 1S
 
\bibitem[Shetye \etal~(2020)]{S20}
{{Shetye}, S., {Van Eck}, S., {Goriely}, S., {Siess}, L., {Jorissen}, A., {Escorza}, A., \& {Van Winckel}, H.} 2020,
\textit{A\&A}, 635L, 6S

\bibitem[Shetye \etal~(2021)]{S21}
{{Shetye}, S., and {Van Eck}, S., {Jorissen}, A., {Goriely}, S., {Siess}, L., {Van Winckel}, H., {Plez}, B., {Godefroid}, M., \& {Wallerstein}, G.} 2021,
\textit{A\&A}, 650A, 118S

\bibitem[Smith \& Lambert (1985)]{smith1}
{{Smith}, V.~V. and {Lambert}, D.~L.} 1985,
\textit{ApJ}, 294, 326S

\bibitem[Smith \& Lambert (1990)]{smith2}
{{Smith}, V.~V. and {Lambert}, D.~L.} 1990,
\textit{ApJ}, 294, 326S

\bibitem[Uttenthaler \etal~(2013)]{utt2013}
{{Uttenthaler}, S.} 2013,
\textit{A\&A}, 556A, 38U

\bibitem[Van Eck \etal~(1999)]{SVE1999}
{{Van Eck}, S. and {Jorissen}, A.} 1999,
\textit{A\&A}, 345, 127V

\bibitem[Van Eck \etal~(2000a)]{SVE2000a}
 {{Van Eck}, S. and {Jorissen}, A. and {Udry}, S. and {Mayor}, M. and {Burki}, G. and {Burnet}, M. \& {Catchpole}, R.} 2000a,
\textit{A\&AS}, 145, 51V

\bibitem[Van Eck \etal~(2000b)]{SVE2000b}
{{Van Eck}, S. and {Jorissen}, A.} 2000b,
\textit{A\&A}, 360, 196V

\bibitem[Van Eck \etal~(2017)]{SVE2017}
{{Van Eck}, S., {Neyskens}, P., {Jorissen}, A., {Plez}, B., {Edvardsson}, B., {Eriksson}, K., {Gustafsson}, B.,  {J{\o}rgensen},  \& {Nordlund}, A.} 2018,
\textit{A\&A}, 601A, 10V


\end{thebibliography}
\end{document}